\documentclass[showpacs,prl,twocolumn,amssymb]{revtex4}
\usepackage{graphicx}

\begin{document}

\title{Critical Zeeman Splitting of Fermi Superfluidity at Infinite Scattering Length}
\author{Lianyi He and Pengfei Zhuang}

\affiliation{Department of Physics, Tsinghua University, Beijing
100084, China}

\begin{abstract}
We determine the critical Zeeman energy splitting for Fermi
superfluidity at infinite s-wave scattering length according to the
Monte Carlo and experimental results of the equations of state.
Based on the universality hypothesis, we show that there exist two
critical fields $H_{c1}$ and $H_{c2}$, between which a
superfluid-normal mixed phase is energetically favored, and
model-independent formulae for $H_{c1}$, $H_{c2}$ and the critical
population imbalance $P_c$ are derived. Using recent Monte Carlo and
experimental results of $P_c$, $H_{c1}$ and $H_{c2}$ are determined.
It is found $H_{c1}=0.41\epsilon_{\text F}$ and
$H_{c2}=0.50\epsilon_{\text F}$, with $\epsilon_{\text F}$ being the
Fermi energy of non-interacting gas.
\end{abstract}

 \maketitle

While Bardeen-Cooper-Schiffer (BCS)
superconductivity/superfluidity in Fermi systems has been
investigted many years ago\cite{BCS}, the main scientific interest
in recent experiments of two-component ultracold Fermi gas is to
create Fermi superfluids in the BCS--Bose-Einstein condensation
(BEC) crossover~\cite{BCSBEC}. At a wide Feshbach resonance point
where the s-wave scattering length $a_s$ diverges, a novel type of
Fermi superfluid has been observed. In the dilute gas limit, there
exists only a single momentum scale in the problem, namely the
Fermi momentum $k_{\text F}=(3\pi^2 n)^{1/3}$ which reflects the
inter-particle distance $n^{-1/3}$. In this so-called unitary
Fermi gas, any physical quantity can be expressed in terms of its
value in the non-interacting case multiplied by a universal
constant~\cite{ho}. For example, the energy density can be written
as ${\cal E}=\xi {\cal E}_0$, where $\xi$ is a universal constant
and ${\cal E}_0$ the energy density of non-interacting Fermi gas.

In addition to the idealized case where fermion pairing happens on
a uniform Fermi surface, the effect of Zeeman energy splitting
$E_Z=\mu_{\text B}H$ between spin-up and -down electrons on BCS
superconductivity was known many years ago~\cite{CC}(in the
following we absorb the magnetic moment $\mu_{\text B}$ in to the
definition of the ``magnetic field" $H$). At a critical Zeeman
field or the so-called Chandrasekhar-Clogston (CC) limit
$H_c=0.707\Delta_0$ where $\Delta_0$ is the zero temperature gap,
a first order phase transition from the gapped BCS state to the
normal state occurs. Further theoretical studies showed that the
inhomogeneous Fulde-Ferrell-Larkin-Ovchinnikov (FFLO)
state~\cite{FFLO} may survive in a narrow window between $H_c$ and
$H_{\text{FFLO}}=0.754\Delta_0$. However, since the thermodynamic
critical field is much smaller than the CC limit due to strong
orbit effect\cite{CC}, it is hard to observe the CC limit and the
FFLO state in ordinary superconductors.

Recent experiments for strongly interacting ultracold Fermi gases
give an alternative way to study the pure Zeeman effect on Fermi
superfluidity~\cite{imbalanceexp}. The atom numbers of the two
lowest hyperfine states of $^6$Li atom, denoted by $N_\uparrow$ and
$N_\downarrow$, are adjusted to create a population imbalance or
polarization
$P=(N_{\uparrow}-N_{\downarrow})/(N_{\uparrow}+N_{\downarrow})$,
which simulates effectively the Zeeman field $H$ in a
superconductor. At unitary, phase separation between unpolarized
superfluid and polarized normal gas, predicted by early theoretical
works~\cite{BCR,Cohen} and Monte Carlo (MC)
simulations~\cite{Carlson}, is observed, but the evidence for the
FFLO and the breached pairing~\cite{BP,he1} states is lacked.

An important scientific problem is to determine the ratio between
the critical Zeeman energy splitting and the Fermi energy for a
homogeneous unitary Fermi gas, $H_c/\epsilon_{\text F}$, which
should be a universal constant at unitary. In an early
work~\cite{Carlson}, the CC limit at unitary was predicted to be
\begin{equation}
\frac{H_c}{\Delta(\mu)}=\frac{1}{\beta}\left(\frac{2^{2/5}}{\xi^{3/5}}-1\right),\label{carl}
\end{equation}
where $\beta=\Delta(\mu)/\mu$ with $\mu$ being the fermion chemical
potential is another universal constant at $H=0$. With the MC data
of $\xi$ and $\beta$, $H_c/\Delta\simeq 1$ is found. However, this
result denotes only the first order superfluid-normal phase
transition point in the grand canonical ensemble with the chemical
potential $\mu$ fixed, and hence not the wanted result
$H_c/\epsilon_{\text F}$. On the other hand, this result is obtained
assuming the normal phase is fully polarized, but recent MC
work~\cite{MC} and experiment\cite{exp} show that the normal phase
at the phase transition is partially polarized. In addition, in
Ref.~\cite{sheehy,he2} it is shown that there exist two critical
fields $H_{c1}$ and $H_{c2}$ in the BCS-BEC crossover and phase
separation is the energetically favored ground state in the region
of $H_{c1}<H<H_{c2}$, like the type-II superconductors. Since the
particle numbers $N_\uparrow$ and $N_\downarrow$ are used as tunable
parameters in MC calculations and experiments, only the first order
phase transition point $(H/\mu)_c$ and the critical spin population
$P_c$ are directly determined, and the two CC limits $H_{c1},H_{c2}$
for homogeneous Fermi gas has not yet been measured in MC
calculations~\cite{MC} and experiments~\cite{exp}.

In this paper, we will determine the two CC limits $H_{c1}$ and
$H_{c2}$ in terms of the Fermi energy $\epsilon_{\text F}$ for a
homogeneous Fermi gas at infinite scattering length, based on the
universal property of the thermodynamics. While a similar approach
was considered in~\cite{forbes,chevy}, the main purpose of our work
is to determine the two critical fields for homogeneous Fermi gas,
and our study should be exact, since we do not assume the normal
phase is fully polarized.

At unitary, we can construct the exact equation of state(EOS) in
the grand canonical ensemble from the universality hypothesis. The
pressure in the polarized normal phase (N) as a function of
averaged chemical potential $\mu=(\mu_\uparrow+\mu_\downarrow)/2$
and the Zeeman field $H=(\mu_\uparrow-\mu_\downarrow)/2$ takes the
form~\cite{ho}
\begin{equation}
{\cal P}_{\text N}(\mu,H)=\frac{2}{5}c\mu^{5/2}{\cal
G}\left(\frac{H}{\mu}\right),\ \ c=\frac{(2M)^{3/2}}{3\pi^2},
\end{equation}
where $\mu_\uparrow$ and $\mu_\downarrow$ are the effective
chemical potentials for the two spin components, $M$ is the
fermion mass. ${\cal G}(x)$ is an unknown universal scaling
function, and only ${\cal G}(0)=\xi_{\text N}^{-3/2}$ is known to
be $\xi_{\text N}\simeq0.56$. The total number density and
magnetization are, respectively, $n_{\text N}(\mu,H)=\partial{\cal
P}_{\text N}/\partial\mu$ and $m_{\text N}(\mu,H)=\partial{\cal
P}_{\text N}/\partial H$. There exists a universal critical value
$\delta_0$ of $H/\mu$, below which the gas is in the partially
polarized state (N$_{\text{PP}}$) with $m_{\text N}<n_{\text N}$.
In the fully polarized normal state (N$_{\text{FP}}$) with
$m_{\text N}=n_{\text N}$, the gas should be non-interacting with
${\cal G}(x)=\frac{1}{2}(1+x)^{5/2}$. While mean field theory
predicts $\delta_0=1$~\cite{sheehy}, it is $3.78$~\cite{MC} from
recent MC simulations.

Monte Carlo studies\cite{MC} and experiments\cite{exp} show that
the stable superfluid phase (SF) at unitary should be unpolarized.
The pressure does not depend on $H$ explicitly, and takes the
well-known form
\begin{equation}
{\cal P}_{\text {SF}}(\mu,H)=\frac{2}{5}c\mu^{5/2}\xi^{-3/2}.
\end{equation}
The total density reads $n_{\text
{SF}}(\mu,H)=c\mu^{3/2}\xi^{-3/2}$.

In the grand canonnical ensemble, a first order SF-N phase
transition occurs when the pressure of the superfluid and the
normal phase equate, i.e., when $H/\mu$ reaches another universal
critical value $\gamma$, which is determined by the algebra
equation
\begin{equation}
\label{gr}
{\cal G}\left(\gamma\right)=\xi^{-3/2}.
\end{equation}
If the normal phase at the phase transition is N$_{\text{FP}}$, we
have $\frac{1}{2}\left(1+H/\mu\right)^{5/2}=\xi^{-3/2}$. Combining
the relation $\Delta(\mu)=\beta\mu$ at unitary, we immediately
recover the result (\ref{carl}) derived in~\cite{Carlson}.

To determine the critical fields $H_{c1},H_{c2}$ for homogeneous
Fermi gas, we turn to the canonical ensemble with fixed total
particle number density $n=(2M\epsilon_{\text F})^{3/2}/(3\pi^2)$.
The Zeeman splitting $H$ is treated as a real external field, and
the conversion between particles in the states $\uparrow$ and
$\downarrow$ is allowed, but the chemical potential $\mu$ is then
not a free parameter.

The chemical potential $\mu_{\text N}(H)$ in the polarized normal
phase is solved from the number equation $n_{\text N}(\mu_{\text
N},h)=n$, and the energy density ${\cal E}_{\text N}(h)=\mu_{\text
N}(H)n-{\cal P}_{\text N}(\mu_{\text N}(H),H)$ reads
\begin{equation}
{\cal E}_{\text N}(H)=\frac{5}{3}\left[\frac{\mu_{\text
N}}{\epsilon_{\text F}}-\frac{2}{5}{\cal
G}\left(\frac{H}{\mu_{\text N}}\right)\left(\frac{\mu_{\text
N}}{\epsilon_{\text F}}\right)^{5/2}\right]{\cal E}_0
\end{equation}
with ${\cal E}_0=\frac{3}{5}c\epsilon_{\text F}^{5/2}$. The
well-known relation ${\cal E}=3{\cal P}/2$~\cite{ho} breaks down
here, since the interacting energy with external field $H$ is
included. In the phase N$_{\text{FP}}$, we have $\mu_{\text
N}(H)=2^{2/3}\epsilon_{\text F}-H$. The
N$_{\text{PP}}$-N$_{\text{FP}}$ transition occurs at
$H_0=2^{2/3}\epsilon_{\text F}\delta_0/(1+\delta_0)$. While mean
field theory predicts $H_0=2^{-1/3}\epsilon_{\text
F}\simeq0.794\epsilon_{\text F}$, we find
$H_0\simeq1.26\epsilon_{\text F}$ from the recent MC simulations.

Solving the number equation $n_{\text{SF}}(\mu,H)=n$ for the
superfluid phase, the chemical potential and energy density are
given by $\mu_{\text{SF}}(H)=\xi\epsilon_{\text F}$ and ${\cal
E}_{\text{SF}}(H)=\xi {\cal E}_0$. At $H=0$, there is the BCS
instability ${\cal E}_{\text{SF}}(0)<{\cal E}_{\text{N}}(0)$ which
is numerically supported by the fact $\xi_{\text N}>\xi$. While
${\cal E}_{\text{SF}}(H)$ keeps independent of $H$, ${\cal
E}_{\text{N}}(H)$ should be a monotonously decreasing function. If
there exists no heterogeneous mixed phase, a phase transition
occurs at ${\cal E}_{\text{N}}(H_c) = {\cal E}_{\text{SF}}$. Once
${\cal G}(x)$ is known, one can determine $H_c$. If we assume the
normal state at $H_c$ is fully polarized, i.e., $H_c\geq H_0$, we
find
\begin{eqnarray}
\frac{H_c}{\epsilon_{\text F}}=2^{2/3}-\xi,\ \ \
\frac{H_c}{\Delta_0}=\frac{1}{\beta}\left(\frac{2^{2/3}}{\xi}-1\right).
\end{eqnarray}
Here $\Delta_0=\beta\xi\epsilon_{\text F}$ is the energy gap at
fixed density. From $H_c\geq H_0$, there is $\xi\leq
2^{2/3}/(1+\delta_0)\simeq0.33$, which is inconsistent with recent
MC result $\xi\simeq0.42$ or $\xi\simeq0.44$. On the other hand,
taking $\xi\simeq0.4-0.5$ and $\beta\simeq1.2$ from the recent MC
data, we have $H_c/\Delta_0>1.5$, which is in contrast with the
constraint $H_c<\Delta_0$~\cite{forbes} to ensure the superfluid
phase is unpolarized. Therefore, the MC data on $\xi$ and $\beta$
indicates that the phase transition is not directly into the state
N$_{\text {FP}}$.

The absence of heterogeneous SF-N mixed phase in above analysis is
not adequate since the first order phase transition is often
associated with the phase separation phenomenon at fixed density.
We then take the heterogeneous mixed phase into account. In this
case, the critical field $H=H_c(\mu)=\gamma\mu$ for the first
order phase transition in the grand canonical ensemble splits into
a lower and a upper critical fields $H_{c1}$ and $H_{c2}$, and the
SF-N mixed phase appears in the region $H_{c1}<H<H_{c2}$. $H_{c1}$
and $H_{c2}$ can be determined by equating the chemical potential
$\mu$ to its value in the superfluid and the normal phase
respectively, $H_{c1}=\gamma\mu_{\text{SF}}$ and
$H_{c2}=\gamma\mu_{\text N}(H_{c2})$, where $\mu_{\text
N}(H_{c2})$ is obtained by the number equation in the normal
state, $\mu_{\text N}(H_{c2})=\xi\epsilon_{\text
F}\left[1-\frac{2}{5}\xi^{3/2}\gamma {\cal
G}^\prime(\gamma)\right]^{-2/3}$. Thus we find the following
model-independent expression for the lower and upper critical
Zeeman fields
\begin{equation}
H_{c1}=\gamma\xi\epsilon_{\text F},\ \ \
H_{c2}=\gamma\xi\epsilon_{\text
F}\left[1-\frac{2}{5}\xi^{3/2}\gamma {\cal
G}^\prime(\gamma)\right]^{-2/3}. \label{CC}
\end{equation}

From the phase equilibrium condition ${\cal
P}_{\text{SF}}(\mu,H)={\cal P}_{\text N}(\mu,H)$, in the mixed
phase the ratio $H/\mu$ keeps a constant $\gamma$ and the chemical
potential reads $\mu_{\text{PS}}(H)=H/\gamma$. The property of the
normal bubble in the mixed phase depends on the value of $\gamma$.
The MC calculations~\cite{MC} predict $\gamma<\delta_0$, i.e., the
normal bubble is partially polarized. The volume fractions of the
superfluid and normal phase, denoted by $x$ and $1-x$
respectively, are determined by
$n=x(H)n_{\text{SF}}(\mu_{\text{PS}},H)+[1-x(H)]n_{\text
N}(\mu_{\text{PS}},H)$. Using the EOS for the phases SF and N, we
find
\begin{equation}
x(H)=\frac{(H/H_{c1})^{-3/2}-(H_{c2}/H_{c1})^{-3/2}}{\frac{2}{5}\xi^{3/2}\gamma
{\cal G}^\prime(\gamma)}.
\end{equation}
The energy density of the mixed phase, ${\cal
E}_{\text{PS}}(H)=\mu_{\text{PS}}n-{\cal
P}_{\text{PS}}(\mu_{\text{PS}},H)$, can be evaluated as
\begin{equation}
{\cal E}_{\text{PS}}(H)
=\frac{5}{3}\frac{H}{H_{c1}}\left[1-\frac{2}{5}\left(\frac{H}{H_{c1}}\right)^{3/2}\right]\xi{\cal
E}_0.
\end{equation}
Since the normal bubble is polarized, there is a nonzero global
polarization $P$ in the mixed phase, i.e., the system becomes
``spontaneously megnetized" when $H>H_{c1}$. From the definition
$P=(N_{\uparrow}-N_{\downarrow})/(N_{\uparrow}+N_{\downarrow})$,
we find
\begin{equation}
P(H)=\gamma^{-1}\left[\left(H/H_{c1}\right)^{3/2}-1\right].
\end{equation}
The mixed phase continuously link the superfluid and normal state.
We have $\mu_{\text{PS}}=\mu_{\text{SF}}$ at $H=H_{c1}$ and
$\mu_{\text{PS}}=\mu_{\text{N}}$ at $H=H_{c2}$, which ensures
$0\leq x\leq 1$ with $x(H_{c1})=1$ and $x(H_{c2})=0$.

There are three constraints on the universal constants and scaling
function. The appearance of the mixed phase requires
$H_{c1}<H_{c2}$, which gives rise to $0<\gamma {\cal
G}^\prime(\gamma)<\frac{5}{2}\xi^{-3/2}$. Since the superfluid
phase is unpolarized, there should be $H_{c1}<\Delta_0$ or
$\beta>\gamma$, which plays the same role as the lower bound for
the ratio $\mu_\uparrow/\mu_\downarrow$ proposed in \cite{forbes}.
Finally, the normal state at $H=H_{c2}$ is partially polarized,
there is $H_{c2}<H_0$.

For the discussions above, the mixed phase is assumed to be the
ground state in the region $H_{c1}<H<H_{c2}$. As a complete study,
we have to prove that the mixed phase has the lowest energy in
this region. While the information on the scaling function ${\cal
G}(x)$ is still lacking, the requirement of the lowest energy can
tell us some constraints on it.

Firstly, the energy density of the mixed phase can be written as
${\cal E}_{\text{PS}}(H)=\frac{5}{3}\xi{\cal E}_0f(H/H_{c1})$ with
$f(z)=z-\frac{2}{5}z^{5/2}$. Since $f^\prime(z)=1-z^{3/2}$, ${\cal
E}_{\text{PS}}(H)$ is a monotonously decreasing function of $H$ in
the region $H_{c1}<H<H_{c2}$. Combining with the fact that ${\cal
E}_{\text{PS}}={\cal E}_{\text{SF}}$ at $H=H_{c1}$ and ${\cal
E}_{\text{SF}}$ is $H$-independent, there is always ${\cal
E}_{\text{PS}}(H)<{\cal E}_{\text{SF}}(H)$ for $H_{c1}<H<H_{c2}$.

\begin{figure}[!htb]
\begin{center}
\includegraphics[width=8cm]{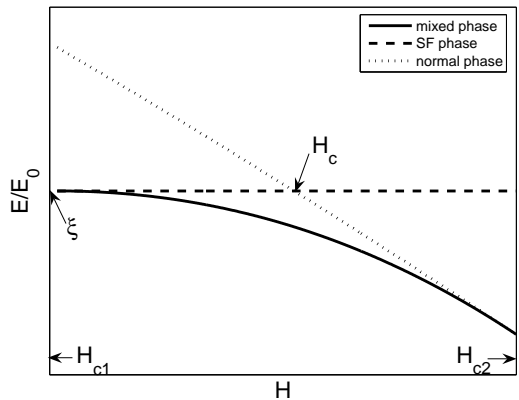}
\caption{The energy for the superfluid, normal and mixed phases in
the region $H_{c1}<H<H_{c2}$. \label{fig1}}
\end{center}
\end{figure}

Secondly, the condition ${\cal E}_{\text{PS}}(H)<{\cal
E}_{\text{N}}(H)$ requires $g(\gamma)<g(\gamma^\prime)$ with
$g(t)=h/t-\frac{2}{5}{\cal G}(t)\left(h/t\right)^{5/2}$,
$h=H/\epsilon_{\text F}$ and $\gamma^\prime=H/\mu_{\text N}(H)$.
Even though the full information of the scaling function ${\cal
G}(x)$ is lacked, it is sufficient to show ${\cal
E}_{\text{PS}}(H)<{\cal E}_{\text{N}}(H)$ at $H\lesssim H_{c2}$,
due to the continuity and the fact of ${\cal
E}_{\text{N}}(0)>{\cal E}_{\text{SF}}(0)$. From the first order
derivative of $g(t)$ at $t=\gamma$,
$g^\prime(\gamma)=\gamma^{-2}h\left[(H/H_{c2})^{3/2}-1\right]$,
$g(t)$ is a decreasing function near $t=\gamma$. Therefore, at
$H\lesssim H_{c2}$ the condition $g(\gamma)<g(\gamma^\prime)$
requires $\gamma^\prime<\gamma$ or $\mu_{\text
N}>\mu_{\text{PS}}$. Since $\mu_{\text N}(0)>\mu_{\text{SF}}$ and
$\mu_{\text{PS}}$ is an increasing function of $H$, we believe the
relation $\mu_{\text{SF}}<\mu_{\text{PS}}<\mu_{\text N}$ holds in
the region $H_{c1}<H<H_{c2}$. A schematic plot of the energy
density for various phases are shown in Fig.\ref{fig1}.

While the universal constant $\xi$ and $\gamma$ has been
determined in MC calculations and experiments, to determine the
upper critical field $H_{c2}$ we need the value ${\cal
G}^\prime(\gamma)$, which is not known so far. However, another
important quantity, the critical population imbalance
$P_c=P(H_{c2})$ is determined\cite{MC,exp}. We now consider
systems with separately fixed $N_\uparrow$ and $N_\downarrow$ and
without conversion between the up and down states. From
$P(H_{c1})=0$ and $P(H_{c2})=P_c$, the ground state is the
unpolarized superfluid at $P=0$ and the SF-N mixed phase for
$0<P<P_c$. In the mixed phase, the effective ``magnetic field" is
given by $H(P)=\gamma\xi\epsilon_{\text F}(1+\gamma P)^{2/3}$, and
the critical population imbalance $P_c$ reads
\begin{equation}
P_c=\frac{2{\cal G}^\prime(\gamma)}{5\xi^{-3/2}-2\gamma {\cal
G}^\prime(\gamma)}. \label{PC}
\end{equation}
It is very interesting that to determine the values of
$H_{c1},H_{c2}$, it is not necessary to know the full imformation of
${\cal G}(x)$, we need only the value of ${\cal G}^\prime(x)$ at
$x=\gamma$. On the other hand, we find there exists a simple
relation between the CC limits in different cases,
\begin{equation}
\frac{H_{c2}}{H_{c1}}=\left(1+\gamma P_c\right)^{3/2}.
\end{equation}

To show the $P_c$ we theoretically obtained above is consistent
with that obtained in MC calculations and experiments, we derive
the energy density ${\cal E}_{\text{PS}}$ as a function of the
ratio $n_{\downarrow}/n_{\uparrow}$\cite{MC}:
\begin{equation}
{\cal E}_{\text{PS}}(n_{\uparrow},n_{\downarrow})=
\frac{3}{5}n_{\uparrow}\frac{(6\pi^2n_{\uparrow})^{2/3}}{2M}I\left(\frac{n_{\downarrow}}{n_{\uparrow}}\right)
\end{equation}
With fixed $N_{\uparrow}$ and $N_{\downarrow}$, the energy density
defined as ${\cal
E}=\mu_{\uparrow}n_\uparrow+\mu_{\downarrow}n_\downarrow-{\cal P}$
satisfies the relation ${\cal E}=3{\cal P}/2$ in all phases, since
$H$ is now no longer treated as an external field. The function
$I(z)$ can be shown to be
\begin{equation}
I(z)=2^{-2/3}\xi\left[(1+\gamma)+(1-\gamma)z\right]^{5/3},
\end{equation}
which is consistent with the formula used in the MC
calculation\cite{MC} to obtain $P_c$.

We can now determine the critical Zeeman fields $H_{c1},H_{c2}$
from the data of $\xi,\gamma$ and $P_c$, and compare them with
that obtained with mean field\cite{sheehy} and beyond mean field
theories\cite{Vei}. The MC and experimental data quite close to
each other. The MC calculation gives $\xi_{\text{MC}}=0.42(1)$,
$\gamma_{\text{MC}}=0.967$, and $P_c^{\text{MC}}=0.389$, while the
experimental data are $\gamma_{\text{EXP}}=0.95$, and
$P_c^{\text{EXP}}=0.36$($\xi$ is not measured in \cite{exp}). We
thus take the MC data to determine $H_{c1}$ and $H_{c2}$.
Substituting them into equation (\ref{PC}), we find ${\cal
G}^\prime(\gamma_{\text{MC}})=2.587$. With equation (\ref{CC}), we
obtain
\begin{equation}
H_{c1}=0.407\epsilon_{\text F},\ \ \ H_{c2}=0.503\epsilon_{\text
F}.
\end{equation}

Our formulae (\ref{CC}) and (\ref{PC}) are model independent. In
mean field theory, we have $\xi_{\text{MF}}=0.5906$ and ${\cal
G}(x)=\frac{1}{2}\left[(1+x)^{5/2}\Theta(1+x)+(1-x)^{5/2}\Theta(1-x)\right]$.
Numerical solution of equation (\ref{gr}) leads to
$\gamma_{\text{MF}}=0.8071$ and ${\cal
G}^\prime(\gamma_{\text{MF}})=2.9307$. Thus we find
$H_{c1}^{\text{MF}}=0.477\epsilon_{\text F}$,
$H_{c2}^{\text{MF}}=0.693\epsilon_{\text F}$ and
$P_c^{\text{MF}}=0.933$, which agree well with the numerical values
obtained in~\cite{sheehy}. One finds the mean field value of
$H_{c2}$ deviates significantly from our result. Also, the result
$H_{c2}=0.693+0.087/N+O(1/N^2)$ obtained by the large-N expansion
method\cite{Vei} is also not consistent with our result.

In summary, we have presented a model independent calculation of
the lower and upper Chandrasekhar-Clogston limits of a unitary
Fermi superfluid. Future studies should focus on the calculation
of the scaling function ${\cal G}(x)$ beyond-mean-field
theories~\cite{Vei,Hu,Son}. Once the universal constant $\xi$ and
function ${\cal G}(x)$ are known, one can directly obtain the
critical polarization $P_c$ from our model independent formula
(\ref{PC}) and check the consistency between theory and experiment
or MC simulation.

The above model independent approach can be generalized to finite
temperature $T$, where both the superfluid and normal phase are
polarized due to thermal excitations. From the universality, the
EOS for the normal phase and superfluid read\cite{ho}
\begin{equation}
{\cal P}_{{\text N},\text{SF}}(T,\mu,H)=\frac{2}{5}c\mu^{5/2}{\cal
G}_{{\text N},\text{SF}}\left(H/\mu,T/\mu\right),
\end{equation}
where we have set $k_{\text B}=1$. The scaling functions for the
superfluid and normal phase should be different.

In the grand canonical ensemble, one expects that the phase
transition along the $T/\mu$ axis is of second order at small
$H/\mu$ and first order at large $H/\mu$. The first order phase
transition is determined by the equation ${\cal G}_{\text
N}\left(H/\mu,T/\mu\right)={\cal
G}_{\text{SF}}\left(H/\mu,T/\mu\right)$, or explicitly
$H/\mu={\cal W}(T/\mu)$ with known ${\cal W}(0)=\gamma$. The first
order phase transition should end at a so-called tricritical point
$(H/\mu,T/\mu)=(a,b)$. At mean field level, it is predicted to be
$(a,b)=(0.70,0.38)$~\cite{parish}.

At fixed total particle number, $\mu$ is not a free variable, and
the tricritical point is characterized by
$(T_{\text{TCP}},H_{\text{TCP}})$. Due to the continuity with the
zero temperature case, for $T<T_{\text{TCP}}$, there exist two
critical fields $H_{c1}(T)=\mu_1{\cal W}(T/\mu_1)$ and
$H_{c2}(T)=\mu_2{\cal W}(T/\mu_2)$ with $\mu_1$ and $\mu_2$ being
the chemical potentials corresponding to $H_{c1}$ and $H_{c2}$.
The mixed phase region $H_{c1}<H<H_{c2}$ should decrease with
increasing $T$, and finally disappears at the tricritical point
with $H_{\text{TCP}}=aT_{\text{TCP}}/b$ and
$\mu_1=\mu_2=T_{\text{TCP}}/b$.

When $N_\uparrow$ and $N_{\downarrow}$ are fixed, for
$T<T_{\text{TCP}}$, the phase separation should be the ground
state in the region $P_1<P<P_2$ with $P_1=P(H_{c1})$ and
$P_2=P(H_{c2})$. At $T\neq 0$, $P_1$ should be nonzero and
increase with temperature, and $P_1=P_2=P_{\text{TCP}}$ at the
tricritical point. Once the scaling function ${\cal G}$ and the
tricritical point $(a,b)$ are known, $P_{\text{TCP}}$ and
$T_{\text{TCP}}$ can be calculated from the following
model-independent formulae:
\begin{eqnarray}
P_{\text{TCP}}&=&\frac{{\cal G}_x^\prime(a,b)}{\frac{5}{2}{\cal
G}(a,b)-a{\cal G}_x^\prime(a,b)-b{\cal
G}_y^\prime(a,b)},\nonumber\\
\frac{T_{\text{TCP}}}{\epsilon_{\text
F}}&=&b\left[\frac{5P_{\text{TCP}}}{2{\cal
G}_x^\prime(a,b)}\right]^{2/3}
\end{eqnarray}
with the definition ${\cal G}_x^\prime(x,y)=\partial{\cal
G}(x,y)/\partial x$ and ${\cal G}_y^\prime(x,y)=\partial{\cal
G}(x,y)/\partial y$, and ${\cal G}$ can be the scaling function of
either the superfluid or the normal phase.

{\bf Acknowledgments:}\ The work is supported by the NSFC Grants
10575058 and 10735040 and the 973-project 2006CB921404.

\end{document}